# The Haute Provence monochromatic heliograph (1958-1994)

Jean-Marie Malherbe

*emeritus astronomer, Paris Observatory-PSL research university, France*

8 February 2024

**Abstract**

Bernard Lyot invented the monochromatic birefringent filter in 1933 in order to investigate the coronal emissions of solar structures above the limb with the coronagraph installed at the Pic du Midi observatory. The filter was improved later and he made the first observations of the chromosphere above the solar disk in 1948, at Meudon. After his death, Grenat and Laborde continued the development in the frame of the coming International Geophysical Year (IGY 1957-1958). A modern Hα heliograph was completed soon and the flare patrol started in 1956. This instrument was reproduced by two companies (SECASI and OPL) and disseminated around the world in order to contribute to the IGY common effort dedicated to the solar activity survey. We describe in this short paper the capabilities of one of these copies operating at Haute Provence station from 1958 to 1994.

**1 - Introduction and context**

Bernard Lyot (1897-1952) invented the coronagraph in order to observe the hot solar corona ($10^6$ K or more) outside eclipses, in clear sky conditions, such as those sometimes available in high mountains. However, observations in white light mainly reveal the electronic corona (Thomson scattering). In order to study coronal structures such as prominences or magnetic loops, emissions of coronal lines such as Hα for prominences, or highly ionized iron for hot loops, must be selected. This requires a narrow bandwidth of a few Angström (Å). Lyot introduced the monochromatic birefringent filter in 1933, and described the developments later, in a big paper (Lyot, 1944). The filter was based on the interference between ordinary and extraordinary rays issued from a birefringent crystal, such as quartz or calcite, providing a channelled spectrum. A Lyot filter is made of several stages, the thickness of stage N being the double of the thickness of stage N-1. Many stages of thickness e, 2e, 4e, 8e… are necessary to make a narrow bandpass filter. The first filters built by Lyot before WW2 were made of quartz (6 stages with e = 2.22 mm), their bandpass was about 3.0 Å. It is difficult to get a narrower bandwidth with quartz, because of its small birefringence, which implies thick crystals (the retardance between rays is proportional to $\Delta n\, e$, where $\Delta n$ is the difference of ordinary and extraordinary refractive indexes, $\Delta n = 0.01$ for quartz). The total thickness of a 6-stage quartz filter was already 140 mm without the polarizers that are necessary between stages ! More selective filters, first with 1.5 Å bandpass, then with 0.75 Å bandpass, where produced by Lyot with the addition of one or two thin calcite stages ($\Delta n = 0.17$).

3.0 Å bandpass filters are sufficient to observe emission lines above the solar limb, but this is not the case for absorption lines of the chromosphere on the disk. For that reason, Lyot tested successfully, in 1948, a 0.75 Å bandpass filter. He unfortunately died a few years later in Egypt, but his collaborators, Henri Grenat and Gérard Laborde, completed the development of the instrument, which was fully operational in 1954 at Meudon (Grenat & Laborde, 1954) and started systematic observations in 1956.

In 1955, Roberts (1957) noticed that the number of birefringent filters in use in the world for the study of the solar chromosphere was more than twenty. It was decided that Meudon will supervise the fabrication of many 0.75 Å bandwidth Hα Lyot filter by the OPL company (Optique et Précision at Levallois, near Paris) as a standard equipment that could be recommended for automatic solar surveys in the frame of IGY 1957 (d'Azambuja, 1959). A collaboration with the SECASI company in Bordeaux (France) was initiated for the construction of automatic heliographs recording the chromospheric activity in Hα, a commercial copy of the original instrument built by Grenat & Laborde (1954). Many heliographs were built by SECASI and disseminated in various observatories, such as Haute Provence (France), Boulder and McMath-Hulbert (USA), Good Hope cape (South Africa), Mitaka (Japan), San Miguel (Argentina) and Purple Mountain (China). These instruments had the advantage of operating automatically; they were able to follow the Sun and used 35 mm cameras (about 2000 images for a 45 m film); there was a timer (one minute cadence typical) and an optical system allowing to incorporate a clock on the pictures. The guiding of the Hα heliograph was done by a small auxiliary white light telescope with four photoelectric cells, keeping the Sun centred and sending signals to the equatorial mount to control the pursuit of the Sun. In 1959, d'Azambuja noticed that 46 instruments worldwide were participating to the international solar survey, globally operating 23.3 hours per day. During IGY (1957-58), 13000 flares were reported. Flare spectrographs were also built, such as the ones at McMath-Hulbert (USA), Pic du Midi (France, Michard et al, 1959), Utrecht (Holland), Crimea (USSR) and Ondrejov (Tchecoslovakia). The patrol was extended to radio wavelengths with radio-telescopes and interferometers in more than twelve countries. Data were reported in the Quarterly Bulletin on Solar Activity (QBSA, there is an on-line archive at Mitaka, Japan, https://solarwww.mtk.nao.ac.jp/en/wdc/qbsa.html, the QBSA was delivered in sixty countries until 2009). Quick informations about flares were dispatched under the form of ursigrams under the auspices of the "Union Radio Scientifique Internationale" (URSI). The daily messages indicated the coordinates of active regions and alerts concerning flare triggering. But let us now come back to the Hα survey and describe the SECASI instrument and the OPL Lyot filter produced for IGY 1957.

**2 - The telescope and the Lyot filter**

This was a 14 cm aperture refractor, 140 cm focal length open at f/10 (Grenat & Laborde, 1954). But there was a magnifying objective (a diverging lens) located at 120 cm of the primary aperture which increased the focal length to 162 cm (f/12) to provide an image of 15 mm diameter on the photographic camera. There was a rotating mechanical shutter and the exposure time was automatically adapted to the seeing conditions with a photoelectric cell, in order to take into account the variations of the sky transparency. The usual exposure time was 1 second. Outside the solar image, there was a gradation of various optical densities for calibration, and a clock showing the date and time in UT. A small fraction of the beam was reflected by a semi-transparent mirror towards an ocular, for optical adjustments and visual inspection of the solar chromosphere.

The pursuit of the Sun was guided by an auxiliary telescope, in white light, of 64 cm focal length. But there was a variable magnifier providing a 14 mm image, whatever the season (the angular diameter of the Sun varies from 31.5 to 32.5 arc minutes between July and January). There was also a circular slit (0.05 mm) around the image; it was divided into four equal quadrants with four photoelectric cells. The electric signals of two opposite cells were compared in order to correct the pointing in right ascension and declination. The precision was of the order of 2 arc seconds.

The original instrument, built by Grenat and Laborde at Meudon, was copied in many exemplars by the SECASI company; one of them was installed at Haute Provence in France (figures 1 and 2).

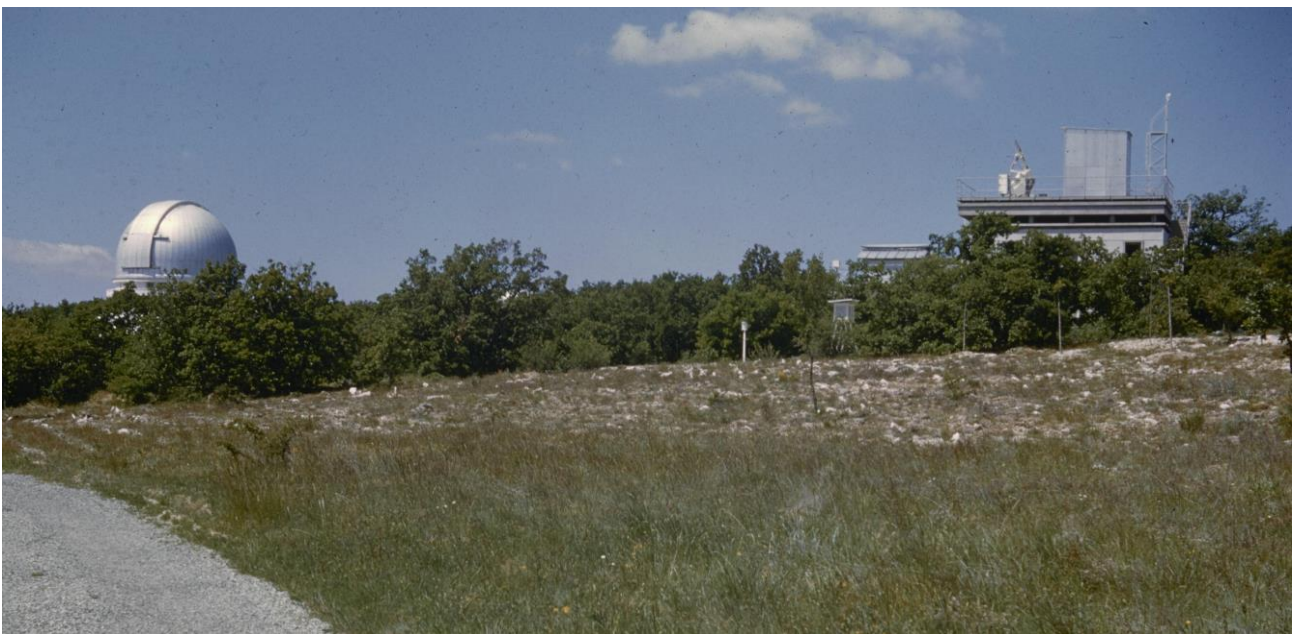

*Figure 1* : the SECASI/OPL heliograph at Haute Provence Observatory at right. Courtesy OP.

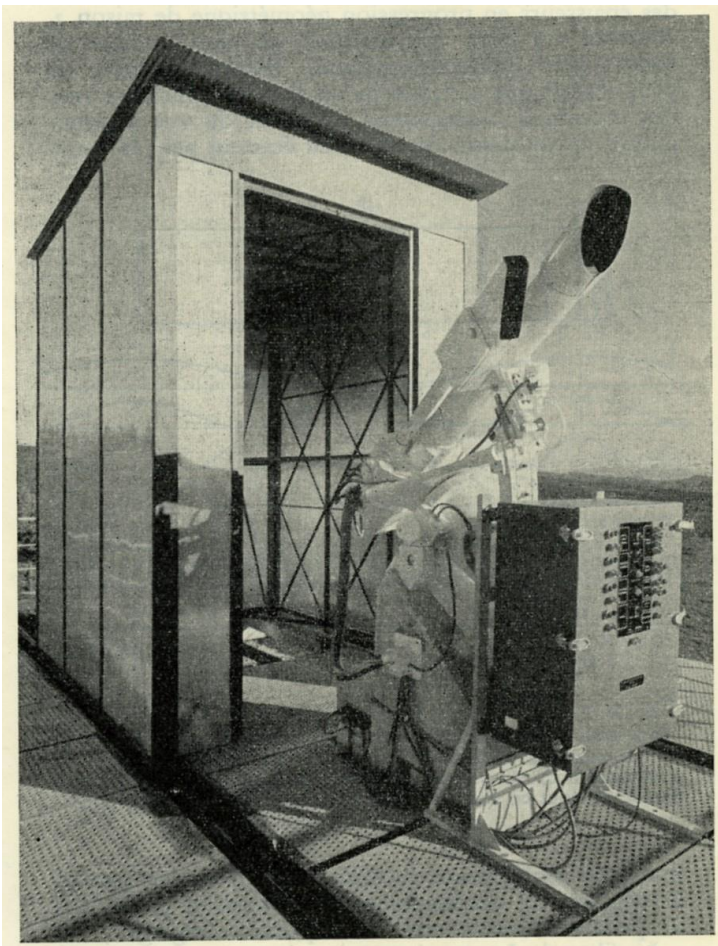

### Figure 2

*The SECASI heliograph at Haute Provence observatory, south of France. It was an industrial copy of the original instrument installed at Meudon. The equatorial mount supported the Hα heliograph and a white light guiding system. It was an automatic instrument. Images of the chromosphere were taken every minute and recorded on 35 mm films.*

FIG. 6. — HÉLIOGRAPHE MONOCHROMATIQUE D'APRÈS LYOT, POUR LA CINÉMATOGRAPHIE DE LA CHROMOSPHÈRE SOLAIRE (MODÈLE SECASI).
(Cliché de l'Observatoire de Haute-Provence.)

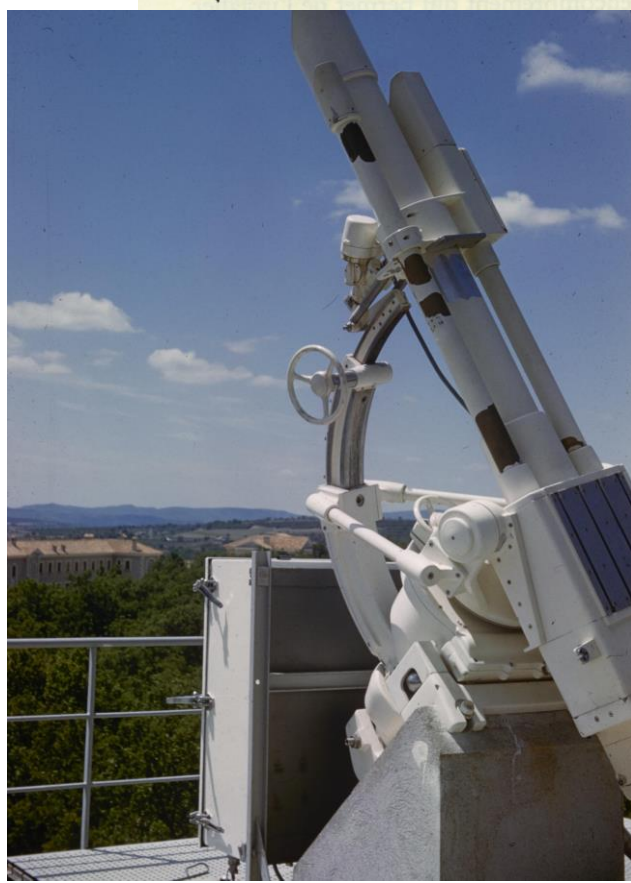
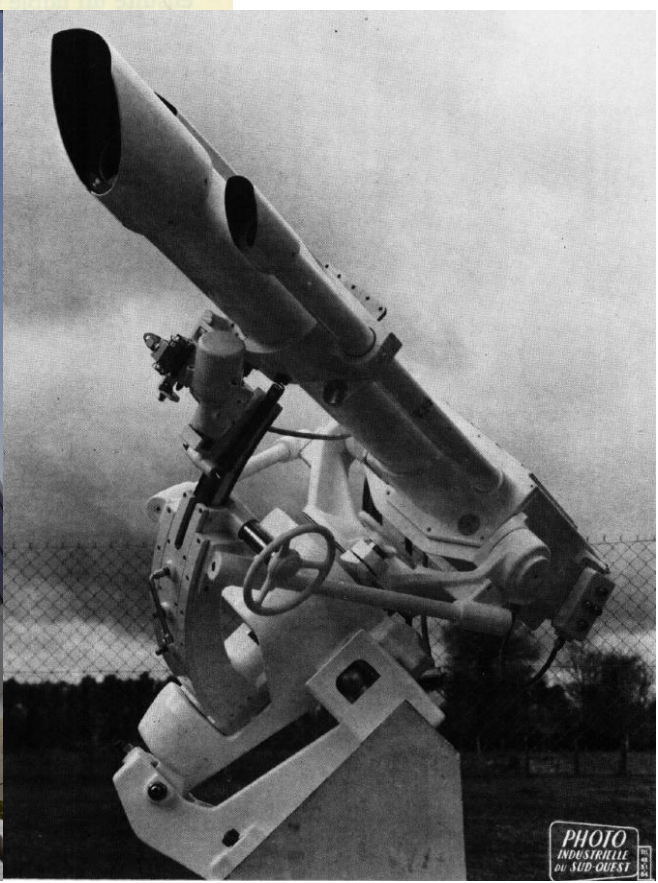

The original Lyot filter, 0.75 Å FWHM, was reproduced by the "Optique et Precision de Levallois" (OPL) company, near Paris, in more than ten exemplars (figure 3). The SECASI heliographs were fitted with the OPL filters. The transmission of the filter is displayed in Figure 4. A red coloured glass filter protected the Lyot monochromator and selected the Hα channel (as Lyot filters deliver a channelled spectrum). The precise tuning of the filter was done by a thermostat; the working temperature was around 40°C and the precision about 0.1°C. The birefringent crystals were enveloped by an electric solenoid and isolating material (cork); the temperature was maintained at the right value by Joule heating and passive cooling.

The film was the 35 mm fine grain Microfile panchro from Kodak, under the form of 45 m reels.

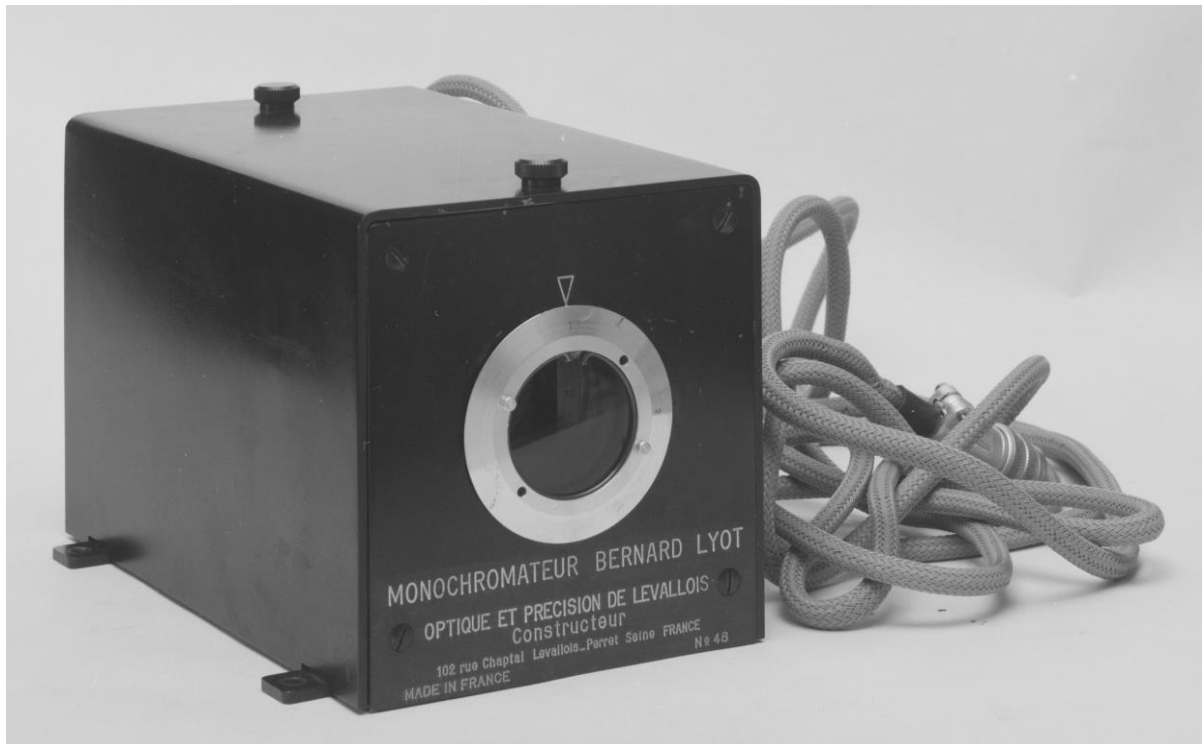

*Figure 3* : the Lyot OPL birefringent monochromatic filter. The electric cable goes to the external thermostat (not shown). Courtesy OP.

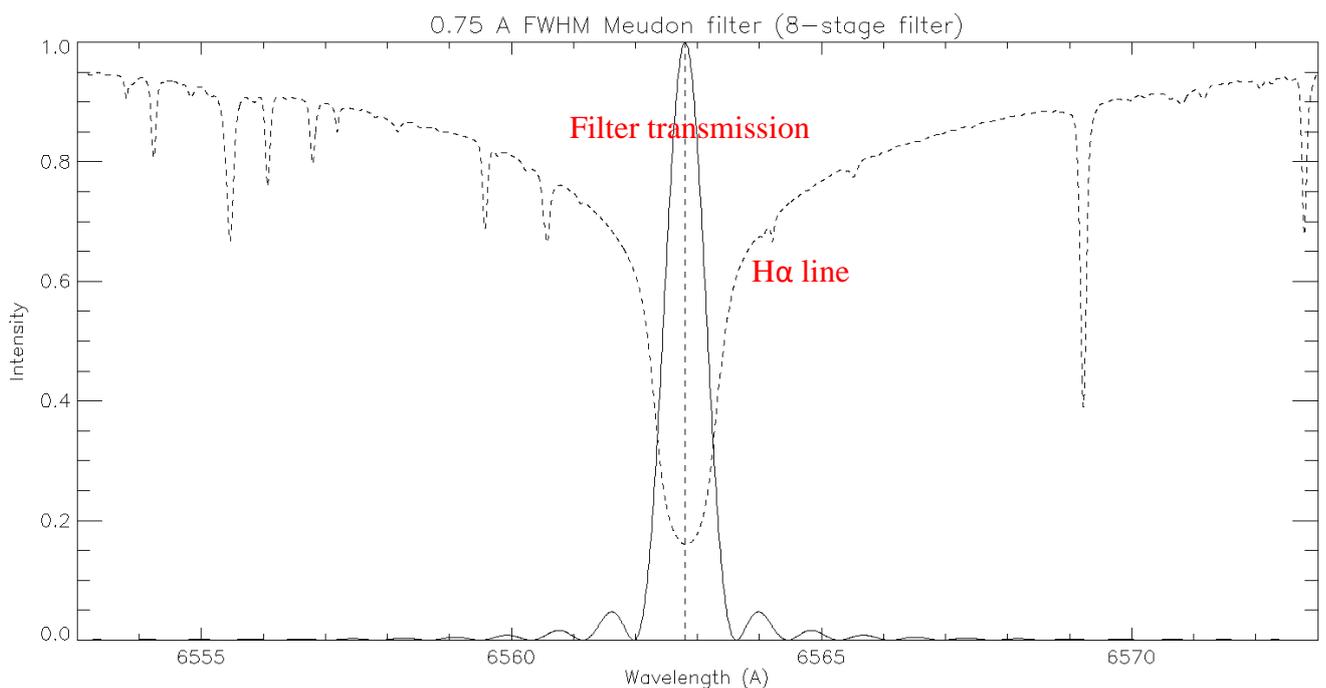

*Figure 4* : the wavelength transmission of the Lyot OPL birefringent monochromatic filter. Courtesy OP.

## 3 - Observations in Haute Provence

The survey of solar flares started in Haute Provence in 1958 and continued on a regular basis until the end of 1994 (figure 5). The routine stopped at the retirement of the observer, André Sanson, who was not replaced. The filter came back regularly at Meudon for a technical visit.

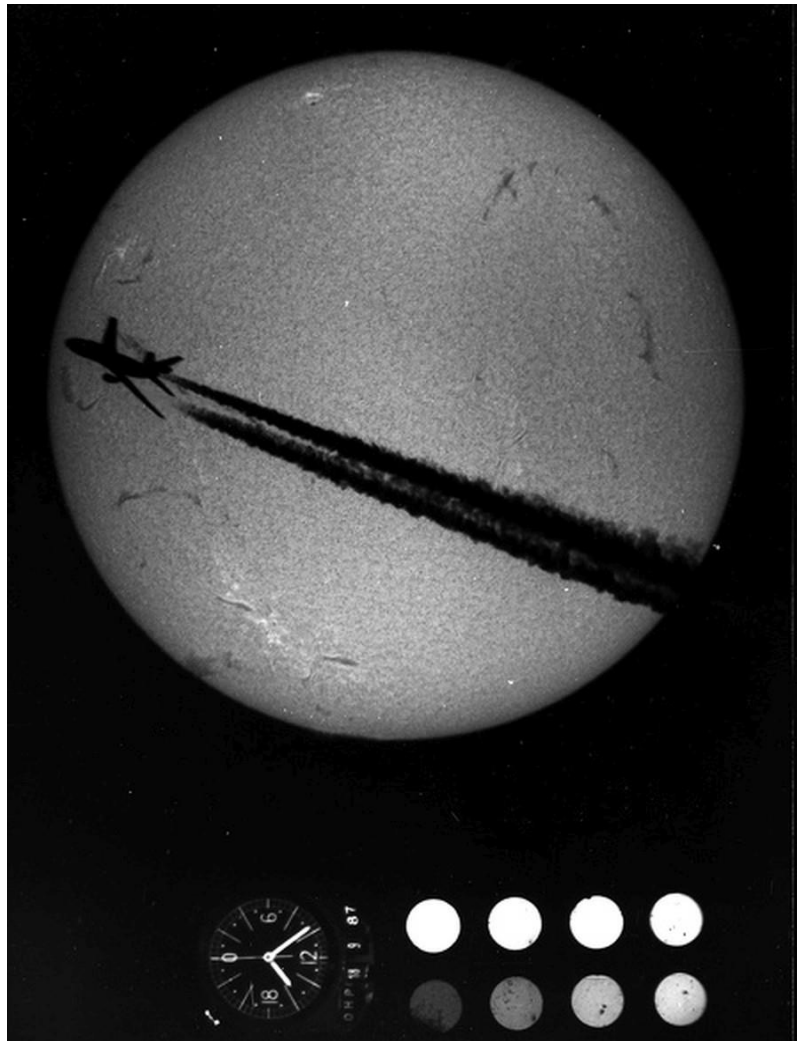

| Year | FILMS (45 m) | Instrument |
|---|---|---|
| 1958 | 14 | Sun diameter 15 mm, 0.75 A FWHM |
| 1959 | 14 | |
| 1960 | 19 | |
| 1961 | 49 | |
| 1962 | 36 | |
| 1963 | 51 | |
| 1964 | 54 | |
| 1965 | 8 | |
| 1967 | 4 | |
| 1968 | 68 | |
| 1969 | 56 | |
| 1970 | 63 | |
| 1971 | 58 | |
| 1972 | 34 | |
| 1974 | 18 | |
| 1975 | 55 | |
| 1976 | 58 | |
| 1977 | 49 | |
| 1978 | 35 | |
| 1979 | 36 | |
| 1980 | 27 | |
| 1981 | 32 | |
| 1982 | 34 | |
| 1983 | 29 | |
| 1984 | 25 | |
| 1985 | 31 | |
| 1986 | 28 | |
| 1987 | 29 | |
| 1988 | 29 | |
| 1989 | 26 | |
| 1990 | 24 | |
| 1991 | 30 | |
| 1992 | 21 | |
| 1993 | 24 | |
| 1994 | 26 | Total 1195 films = 54 km |
| 1995 | 1 | or 2 200 000 images |

*Figure 5* : *the annual number of Hα films (45 m length) produced by the Haute provence heliograph and a typical image, showing the solar chromosphere, the clock for date and time (UT), and the set of neutral densities for photographic calibration. Courtesy OP.*

About 1200 movies, corresponding to 2200000 images, where taken by the Haute Provence heliograph. The movies consist of normal exposures for the disk, revealing the chromospheric activity (filament instabilities or "Disparitions Brusques", flares, starting points of coronal mass ejections), and sometimes a long exposure picture for solar prominences at the limb (in that case, the disk is burned out, because prominences are 5-10 times fainter). The films are archived at the solar service of Meudon observatory.

Two examples of movies are available under the links provided by Figure 6 (first movie, 1958) and Figure 7 (last movie, 1994).

Thanks to Isabelle Bualé, a catalog of observations (PDF format) is available, and the movies can be digitized with high resolution (1 arc second/pixel or better) upon request:

https://www.lesia.obspm.fr/perso/jean-marie-malherbe/OHP/heliograph_OHP.pdf

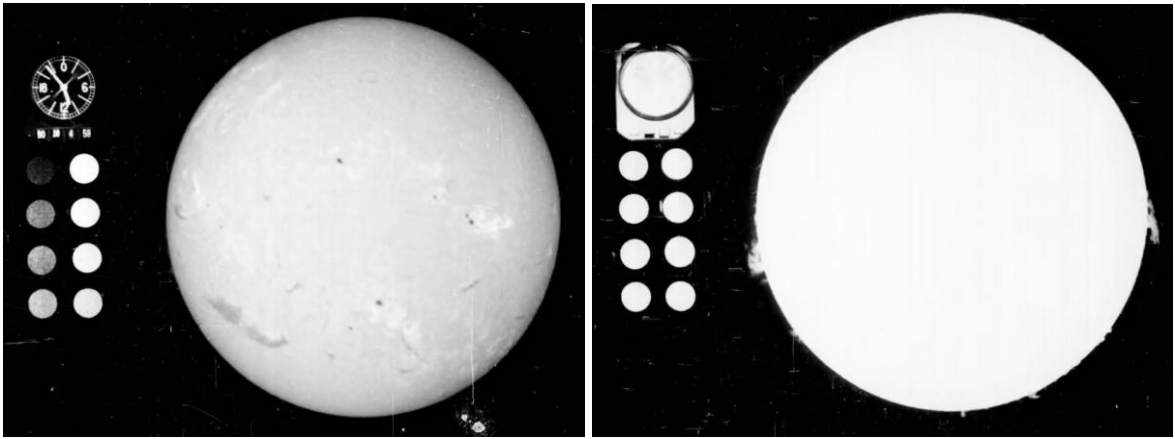

*Figure 6* : first Hα movie produced by the Haute Provence heliograph (left: normal exposure; right: long exposure). Courtesy OP. Link to the movie:

https://www.lesia.obspm.fr/perso/jean-marie-malherbe/OHP/1958-1.mp4

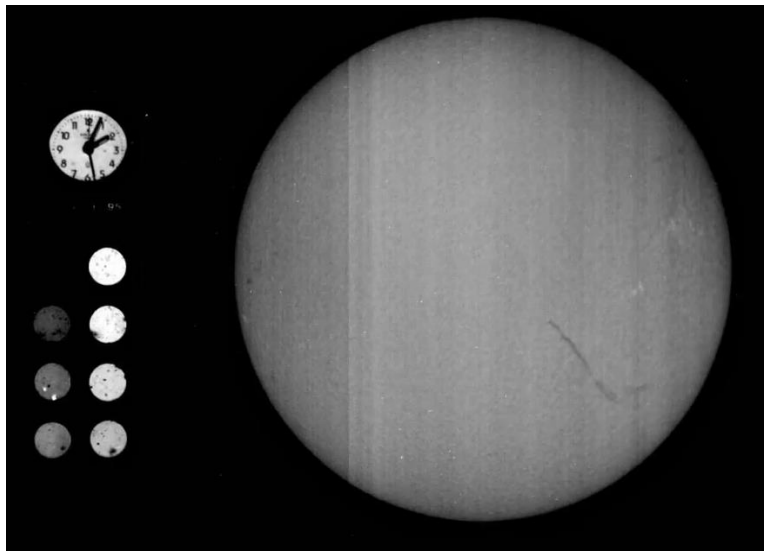

*Figure 7* : last Hα movie produced by the Haute Provence heliograph. Courtesy OP. Link to the movie:

https://www.lesia.obspm.fr/perso/jean-marie-malherbe/OHP/1994-27.mp4

## 4 – References